\documentclass[%
 aip,
 amsmath,amssymb,
 reprint,%
]{revtex4-1}

\usepackage{graphicx}% Include figure files
\usepackage{dcolumn}% Align table columns on decimal point
\usepackage{bm}% bold math
\usepackage[utf8]{inputenc}
\usepackage[T1]{fontenc}
\usepackage{mathptmx}
\usepackage{etoolbox}

\makeatletter
\def\@email#1#2{%
 \endgroup
 \patchcmd{\titleblock@produce}
  {\frontmatter@RRAPformat}
  {\frontmatter@RRAPformat{\produce@RRAP{*#1\href{mailto:#2}{#2}}}\frontmatter@RRAPformat}
  {}{}
}%
\makeatother
\begin{document}

\preprint{AIP/123-QED}

\title{Impact of Au Ion Implantation on 2D $Cr_2Ge_2Te_6$ for Spintronics}
% Force line breaks with \\
\author{Gurupada Ghorai}
\affiliation{School of Physical Sciences, National Institute of Science Education and Research (NISER) Bhubaneswar, An OCC of Homi Bhabha National Institute, Jatni-752050, Odisha, India.}%Lines break automatically or can be forced with \\
\author{Kalyan Ghosh}
\affiliation{School of Physical Sciences, National Institute of Science Education and Research (NISER) Bhubaneswar, An OCC of Homi Bhabha National Institute, Jatni-752050, Odisha, India.}%Lines break automatically or can be forced with \\
\author{Pratap K. Sahoo}%
\email[Correspondence author: ]{pratap.sahoo@niser.ac.in}
\affiliation{School of Physical Sciences, National Institute of Science Education and Research (NISER) Bhubaneswar, An OCC of Homi Bhabha National Institute, Jatni-752050, Odisha, India.}%

\affiliation{Center for Interdisciplinary Sciences (CIS), NISER Bhubaneswar, Jatni-752050, Odisha, India.}%

\date{\today}% It is always \today, today,

\begin{abstract}

Advancements in 2D magnetic materials highlight their potential in semiconductors, magnetism, and spintronics, particularly in tuning magnetic properties for spintronic applications. This study investigates the impact of low-energy (30 KeV) Au ion implantation on 2D layered exfoliated $Cr_2Ge_2Te_6$ flakes prepared on Si/SiO$_2$ substrates using the Scotch tape method. Five different ion doses (5$\times10^{13}$, 1$\times10^{14}$, 5$\times10^{14}$, 1$\times10^{15}$, and 2.5$\times10^{15}$ ions/cm$^2$) were used to modify the morphology, composition, structural, and vibrational properties of the samples. The implantation introduces significant changes in morphology and magnetic behavior, leading to an increase in Curie temperature and an attribution from superexchange to double exchange interactions. The reduced exchange energy gaps and modified magnetic moments attribute to Au ions intercalation in $Cr_2Ge_2Te_6$ underscore the potential of ions implantation to tune the magnetic properties of 2D materials for advanced spintronic applications.

\begin{description}
\item[Usage]
30 keV Au ion beam accelerator, FESEM, EDX, XRD, Raman spectroscopy, TEM, SRIM\\ and TRIDYN software packages, SQUID magnetometry.
\item[Structure]
Structural modification, Ion beam defect analysis, magnetization measurement. 
\end{description}
\end{abstract}
%\keywords{Suggested keywords}%Use showkeys class option if keyword                            %display desired
\maketitle
%\tableofcontents
\section{\textbf {Introduction}}

The research field focusing on 2D magnetic materials has witnessed remarkable advancements in various areas such as semiconductors, magnetism, medicine, and spintronics\cite{Manzeli2017,doi:10.1126/science.aac9439,doi:10.1073/pnas.0502848102,doi:10.1021/acsnano.9b03632}. These materials have shown promise for future applications in supercomputers and memory devices\cite{doi:10.1021/acsaelm.2c00419,9,10}. In the context of spintronics and magnetic applications, 2D ferromagnetic materials are particularly intriguing, as they offer the possibility of harnessing the spin degrees of freedom of electrons to enhance magnetic functionality. It is important to understand the spin texture in these materials to realize their potential in device applications. The Mermin-Wagner theorem, a fundamental result in statistical physics, states that in the two-dimensional isotropic Heisenberg model, magnetic order is prohibited at any finite temperature due to thermal fluctuations and entropy\cite{PhysRevLett.17.1133}. This theorem suggests that 2D ferromagnetism is not possible in isotropic systems. However, Cheng Gong et al. proposed the Ising model, which allows for stable 2D ferromagnetism by removing the restrictions imposed by the Mermin-Wagner theorem\cite{Gong}. The Ising model considers anisotropic interactions between spins and provides a framework for understanding and predicting ferromagnetic behavior in 2D systems. The discovery of stable 2D ferromagnetism has opened up new possibilities for exploring their magnetic properties and developing applications in the field of spintronics. Researchers continue to investigate and characterize these materials to discover their unique properties and potential for future technological advancements.

Achieving stable 2D long-range ferromagnetic (FM) or antiferromagnetic (AFM) order with high transition temperature in the presence of thermal fluctuations is a fascinating and challenging research area of 2D phase-change materials. The superexchange and direct-exchange interactions control the FM order in 2D magnetic materials. The superexchange interaction, which involves a 90-degree cation-anion-cation configuration, plays a significant role in determining the magnetic properties\cite{Kobayashi2016}. The distance between the cation and anion atoms is crucial because the 90-degree interaction dominates over the 180-degree interaction that involves the cation-cation interaction. Another interesting property of 2D material flakes is their easy exfoliation nature, keeping the intrinsic magnetism intact, making them a candidate for spin-based devices such as FET transistors, spintronic memory, spin-logic devices, etc.\cite{10.1088/0022-3727/46/7/074003, 10.1038/nnano.2010.31}. Layer-dependent materials such as $CrXTe_3$ (where X = Si, Ge, and Sn) exhibit different magnetic behaviors, including antiferromagnetism or ferromagnetism, depending on the lattice structure and the competition between superexchange and direct exchange interactions\cite{Huang2018}. Among them, $Cr_2Ge_3Te_6$(CGT) is a special 2D magnet with perpendicular magnetic anisotropy and gate-controllable properties below its Curie temperature ($T_C$), which leads to several optoelectronic and nanoelectronic devices. Zhenqi Hao et al. explained the structural properties of CGT using scanning tunneling microscopy, and its electronic properties, such as the bandgap, were calculated using density functional theory (DFT) calculations\cite{Gong, Zhenqi}. Doping and intercalating with transition-metal ions can tune the magnetic, electrical, and optical properties of the 2D materials. In addition, introducing vacancies and defects strongly reduces the formation energy of magnetic elements that assist in manipulating ferromagnetism in 2D materials. Ion beam implantation is a widely used technique for modifying and enhancing material properties for specific applications. By employing low-energy ion beams in the KeV range, controlled doping and the introduction of functional defects, such as point defects, extended defects, and interstitial defects, can be introduced close to the surface\cite{PhysRevLett.110.015501, Gasiorek, MOHAPATRA2020106504}. This technique offers opportunities for tailoring the magnetic and electronic properties of 2D materials, further expanding their potential applications by modifying the exchange, superexchange, and double exchange interactions through controlled doping and introducing specific defects. Various studies have utilized ion beam implantation to modify the properties of materials and achieve desired magnetic behaviors \cite{10.1007/978-3-540-45298-0, SAHOO2005188, doi:10.1063/5.0010948}. Introducing defects into $Cr_2Ge_2Te_6$ to tailor its FM properties by ion implantation has not been explored yet. Introducing controlled defects into CGT materials through low-energy ion implantation offers a promising approach for customizing their FM properties, such as transition temperature and exchange energy gap, which are crucial for optimizing spintronic applications. 

Here, we focus on defects assisting magnetic, vibrational, and optical properties in thin layers of 2D-CGT. The bulk sample of CGT was prepared by annealing the proportionate ratio of elemental powders in an ample through various steps. To obtain 2D nanoscale flakes, we used the Scotch tape method to transfer the material onto Si/SiO$_2$ substrates. 30 keV Au ions was used to introduce controlled defects on the layers of flakes. The implanted samples were then characterized to obtain structural and morphological information as a function of ion doses. Raman spectroscopy was used to investigate the lattice vibrations of the material. The temperature-dependent magnetic measurement as a function of defect concentration by ion implantation explored the induced transition temperature and reduction of the exchange energy gap of CGT.
\section{Experimental Methods}
The process of synthesizing the CGT material involved several steps: high-purity (99.995\%) commercial materials of Cr, Ge, and Te were mixed in a molar ratio of 1:2:6. This mixing step ensured the appropriate composition of the material. The mixed sample materials were kept in a vacuum sealed ample and annealed gradually at a rate of 5$^\circ$C/min till 1050$^\circ$C and then kept for eight days. This annealing process allowed the formation and stabilization of the desired crystalline structure. After the annealing process, the sample was cooled down at a rate of 10$^\circ$C/hour and maintained at 800$^\circ$C for four days before cooling down to room temperature. This controlled cooling process helped to prevent rapid changes in the material's structure and allowed it to reach a stable state. After obtaining the bulk sample of CGT, the composition was checked using Energy Dispersive X-ray Spectroscopy (EDX). The thin 2D layers of CGT flakes were transferred onto Si/SiO$_2$ substrates using the Scotch tape method, providing a suitable platform for further characterization. 

Five samples of 2D layered CGT with similar conditions were prepared on Si/SiO$_2$ substrates for ion implantation. 30 kV Ion Accelerator Facility in the low energy ion beam facility of the Institute of Physics (IOP), Bhubaneswar, India, has been utilized for performing the Au ion implantation experiments under a vacuum of 1$\times10^{-7}$ mbar. Low energy Au ions were implanted into the 2D flakes at five different doses of 5$\times10^{13}$, 1$\times10^{14}$, 5$\times10^{14}$, 1$\times10^{15}$, and 2.5$\times10^{15}$ $ions/cm^2$. Characterization techniques such as Field Emission Scanning Electron Microscopy (FESEM)  and EDX mapping were employed to analyze the morphology and composition of the samples. Structural information was obtained using Grazing Incidence X-ray Diffraction (GIXRD) at a fixed incident angle $1^{\circ}$ analysis with a Rigaku system. The microscopic crystallographic information after ion implantation was studied by cross-sectional transmission electron microscopy (XTEM). The lamella for XTEM was prepared using a 30 keV Ga source-based dual beam focused ion beam (FIB) system. Confocal Raman spectroscopy was used to study the effect of ion fluence on the lattice vibration modes of the samples. Furthermore, the DC magnetization properties of the samples were investigated using a vibrating sample magnetometer in a SQUID system (MPMS-3, Quantum Design).
\section{Results and discussions} 
\begin {figure}[!ht]
\centering
\includegraphics[width=3.4in]{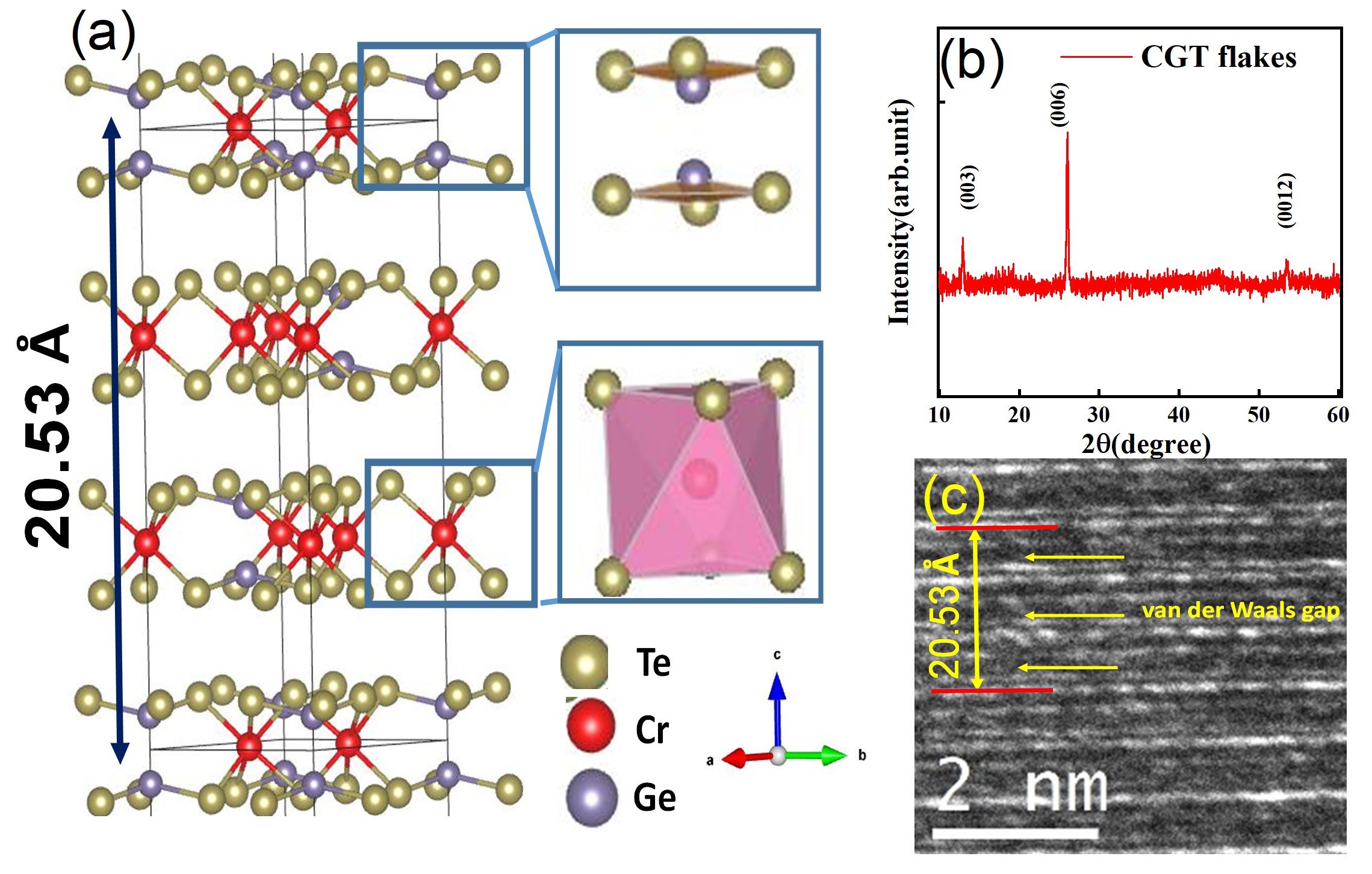}
\caption{(a) Schematic illustration of crystal structure of CGT. (b) GIXRD of pristine samples. (c) Cross sectional TEM image of CGT.}
\label{schematic}
\end{figure}
The CGT is a 2D FM insulator with a $T_C$ of $\approx$ 62 K and an insulator below 100K\cite{Zhang_2016}. The crystal structure of crystalline CGT contains three atomic slabs with three van der Waals gaps in an ABC sequence along the c-axis in a hexagonal unit cell. For each atomic slab, one Cr atom forms an octahedral pattern with six Te atoms, and the two Ge atoms form intergrown tetrahedra with the six Te atoms present. The position of the Cr atom at the centers with slightly distorted octahedra Te atoms and the two Ge atoms forms Ge$_2$Te$_6$  ethane-like molecule. The grown crystals are flake-like structures of 2-3 mm in size. Figure \ref{schematic}(a) shows the crystal structure of CGT with sub-unit centered at Cr and Ge sites. The typical XRD pattern of a highly crystalline structure has (003), (006), and (0012) crystal planes, as shown in Figure \ref{schematic}(b). The identification of the layer structures is also done by atomic resolution TEM imaging shown in Figure \ref{schematic}(c). 

\begin {figure}[!ht]
\centering
\includegraphics[width=3.4in]{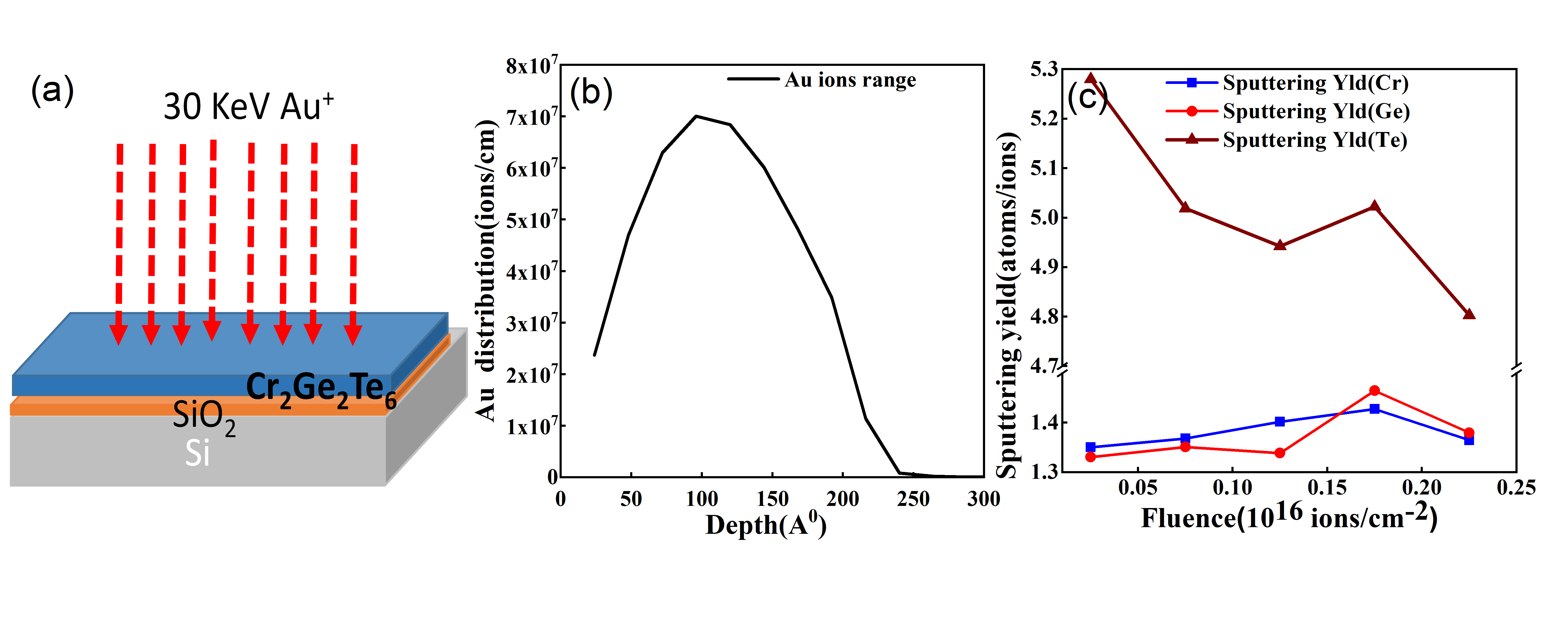}
\caption{(a) Schematic illustration of 30 KeV Au ion implantation on CGT/SiO$_2$/Si substrate. (b) 30 KeV energy Au ion stopping range in the CGT 2D flakes calculated from the SRIM calculation; (c) Sputtering yield of the Cr, Ge, and Te are calculated using TRIDYN software.}
\label{SRIM}
\end{figure}
30 keV Au ion implantations studied the defect engineering on magnetism and vibrational modes. Figure \ref{SRIM}(a) shows the schematic of ion implantation in CGT on the effect of using ions beam techniques and simulation software to study thin films. In the case of 30 KeV Au ion implantation on Si/SiO$_2$ substrates, the SRIM simulation was performed to know the distribution of Au ions in the CGT as a function of depth. From Figure \ref{SRIM}(b), it is clear that the 30 keV Au can penetrate $\approx$ 20 nm with a projected range (R$_p$) of 12 nm. It should be noted that the nuclear energy loss ($S_n$=27.6 eV/nm) is one order of magnitude higher than the electronic energy loss ($S_e$=2.2 eV/nm) \cite{Toulemonde}. This suggests that low energy ion implantation can create various defects, including bond breaking and surface sputtering \cite{doi:10.1063/5.0027462}. The TRIDYN simulation was performed to determine the sputtering yield of different elements (Cr, Ge, and Te) during ion implantation, as shown in Figure \ref{SRIM}(c). The sputtering yield represents the number of atoms sputtered out from the material per incident ion. In this case, the sputtering yield of Ge increases with fluence, indicating that more Ge atoms are sputtered out with ions fluence. On the other hand, the sputtering yield of Te decreases with fluence because of the weak van der Waals bonding. In contrast, the sputtering yield of Cr is around 1.2 atoms/ions and almost slightly increases with fluences. The later part will correlate these results with other experimental observations to understand the effect of ion-induced defects.

\begin {figure}[!ht]
\centering
\includegraphics[width=3.4in]{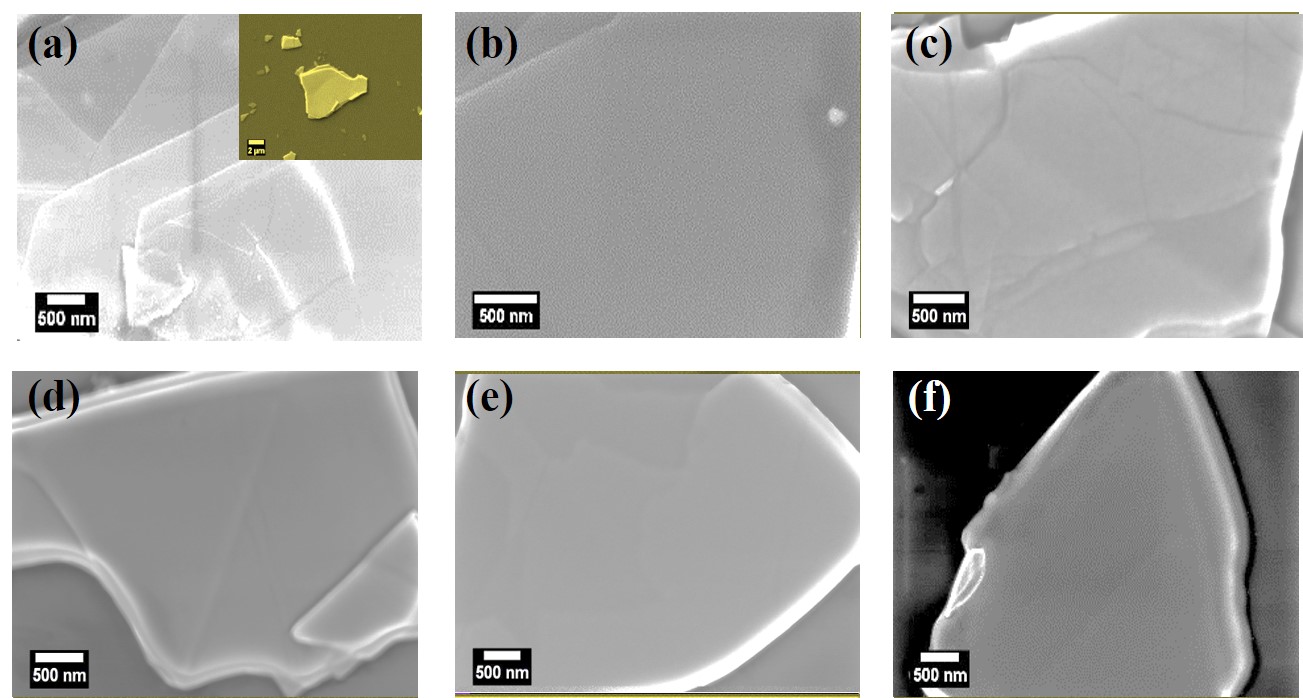}
\caption{FESEM and Surface false color morphology of the pristine and implanted sample are present here. (a) Pristine, (b) 5$\times10^{13}$, (c) 1$\times10^{14}$, (d) 5$\times10^{14}$, (e) 1$\times10^{15}$, (f) 2.5$\times10^{15} ions/cm^2$}
\label{FESEM}
\end{figure}

The ion species, energy, and fluence can significantly affect the surface morphology during ion implantation. Figure \ref{FESEM} shows the FESEM image of pristine and implanted samples with various doses of ion. The pristine sample (Figure \ref{FESEM} (a)) shows that flakes are not monolayers, there are stacking of few layers. But after ion implantation, there are no significant changes observed from FESEM images up to a fluence of 1$\times10^{14} ions/cm^2$ in Figure \ref{FESEM} (b) and (c). With increasing fluence up to 2.5$\times10^{15} ions/cm^2$, the edge and surface of the CGT flake look like melting due to high fluence implantation as shown in Figure \ref{FESEM} (d-f). This observation of surface melting is the effect of the thermal spike model, which arises due to the combination of $S_n$ and $S_e$, which creates high temperature via high electron-phonon coupling within the ion track on a short time scale. As the ion fluence increases, the overlapping region of the ion beams is larger, and the thermal spike effect is higher. The results obtained from EDX analysis typically represent the percentage of composition for each element detected, shown in Table \ref{EDS percenatge}. It should be noted that the sputtering yield removes the surface atoms. According to the composition percentages, there is a noticeable reduction in the atomic percentage of Cr, Ge, and Te in the samples implanted with increasing doses of ion that corroborate with the TRIDYN simulation. The atomic percentage of Au increases with ion fluence because the Au atoms are incorporated inside the top 20 nm of CGT layers. The atomic percentage modification caused by ion implantation and the changes in composition can have significant implications for spin reorientation within the lattice.

\begin{table}[!htbp]
\caption{\label{sec:level21} Atomic composition percentage table from the EDX mapping of all samples at the edges of CrK, GeK, TeL, and AuL. }
\label{EDS percenatge}
\centering
\begin{tabular}{|c|cccc|}
\hline
Sample&Cr$_K$& Ge$_K$&Te$_L$&Au$_L$\\\hline
unit (Atomic composition)&\%&\%&\%&\%\\\hline
Pristine &\mbox{22.3}&\mbox{17.5}&\mbox{60.2}&\mbox{0.0}\\
5$\times10^{13}$&\mbox{28.5}&\mbox{15.2}&\mbox{56.2}&\mbox{0.2}\\
1$\times10^{14}$&\mbox{28.4}&\mbox{14.8}&\mbox{56.1}&\mbox{0.2}\\
5$\times10^{14}$&\mbox{30.1}&\mbox{13.3}&\mbox{56.3}&\mbox{0.3}\\
1$\times10^{15}$&\mbox{31.1}&\mbox{12.6}&\mbox{55.6}&\mbox{0.8}\\
2.5$\times10^{15}$&\mbox{31.2}&\mbox{12.5}&\mbox{55.2}&\mbox{1.1}\\
\hline
\end{tabular}
\end{table}
\begin {figure}[!ht]
\centering
\includegraphics[width=3.4in]{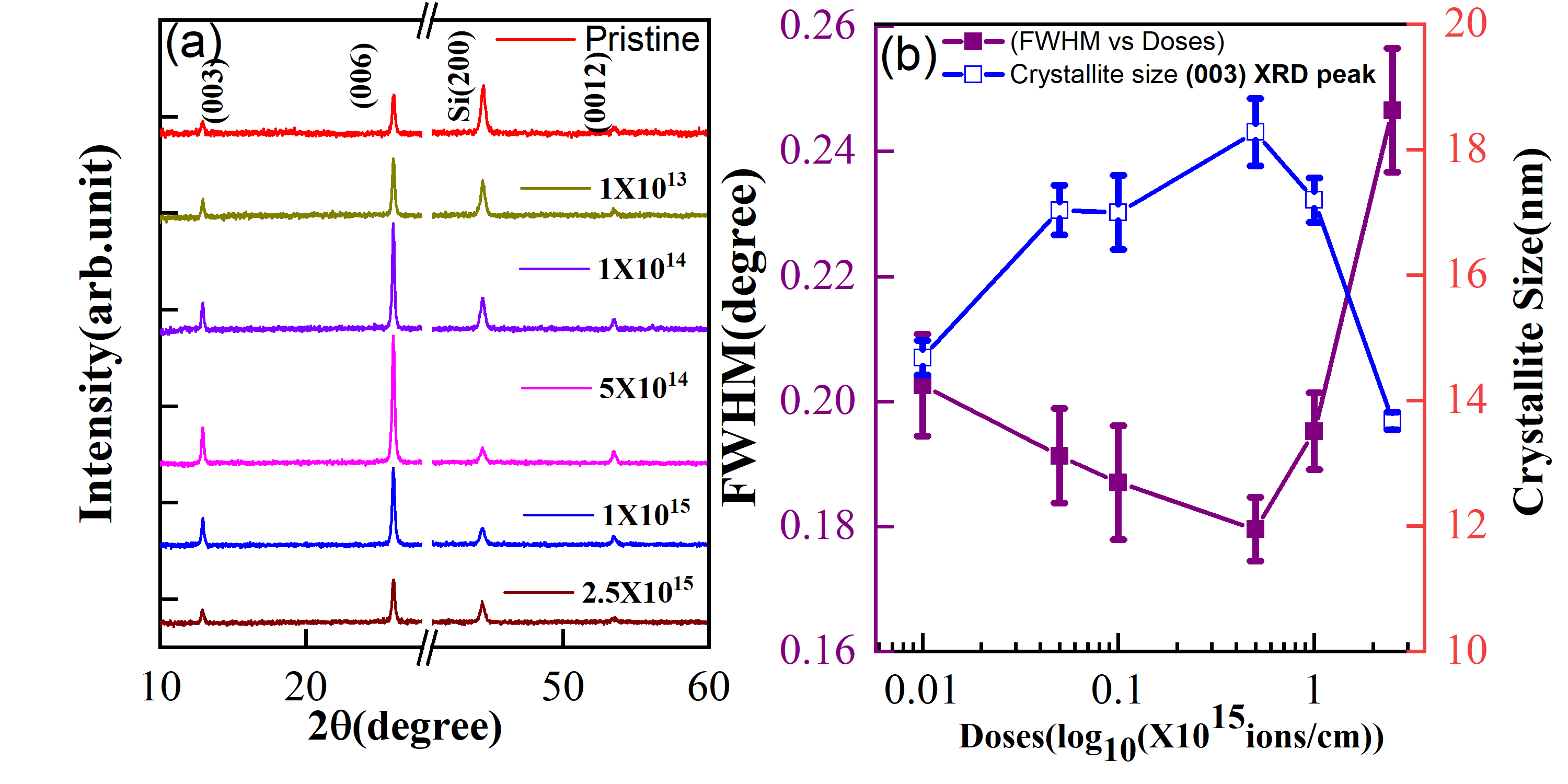}
\caption{(a) GIXRD of pristine and implanted samples. (b) The calculated FWHM and crystallite size from the GIXRD peaks with the function of ion doses.}
\label{XRD}
\end{figure}

In order to obtain the quality of crystal growth and orientation of crystallites, GIXRD was performed in the 2$\theta$  range of 20-60$^\circ$. Figure \ref{XRD}(a) shows the GIXRD patterns of the pristine and implanted samples 2D CGT at room temperature. The GIXRD peaks are observed in 2$\theta$ values at 12.96, 26.00, and 53.62, corresponding to (003), (006), and (0012) crystal planes, respectively. Along with that, the peak observed at 44.52$^0$ in Figure \ref{XRD}(a) corresponds to the (002) plane of Si. This indicates the single crystalline growth of CGT with (00$l$), and the crystal surface is normal to the c-axis of the flakes. This crystal structure was previously reported by Gong et al.\cite{Gong}, and by analyzing the XRD peaks, the lattice parameters of the CGT crystal were determined to be a=b=6.82 \AA, c=20.53 \AA, and the angles $\alpha=\beta=90$, $\gamma=120$. The GIXRD pattern of the pristine and implanted samples show characteristic peaks corresponding to a single trigonal crystal structure with the space group R$\bar3$ [148]. Figure \ref{XRD}(b) shows the FWHM and crystallite size of the GIXRD peak as a function of ion fluences. The crystallite size is calculated using the Scherrer Equation $L=K\lambda/(\beta cos(\theta))$. Where L defines the nano crystallite size, $\lambda$ is the wavelength of incident X-rays, and $\beta$ is FWHM of peaks at any $2\theta$ position. Initially, the FWHM of the (006) peak remains relatively flat, indicating an unchanged crystalline quality or an unmodified structural order caused by low fluence ion implantation. However, at higher ion fluences, the FWHM of the (006) peak starts to increase. The modified FWHM indicates a more significant impact on the lattice and potentially causing damage or disruption of the crystalline structure. The wider XRD peak suggests a decrease in crystalline quality or an increase in structural disorder owing to higher ion fluences.

\begin {figure}[!ht]
\centering
\includegraphics[width=3.4in]{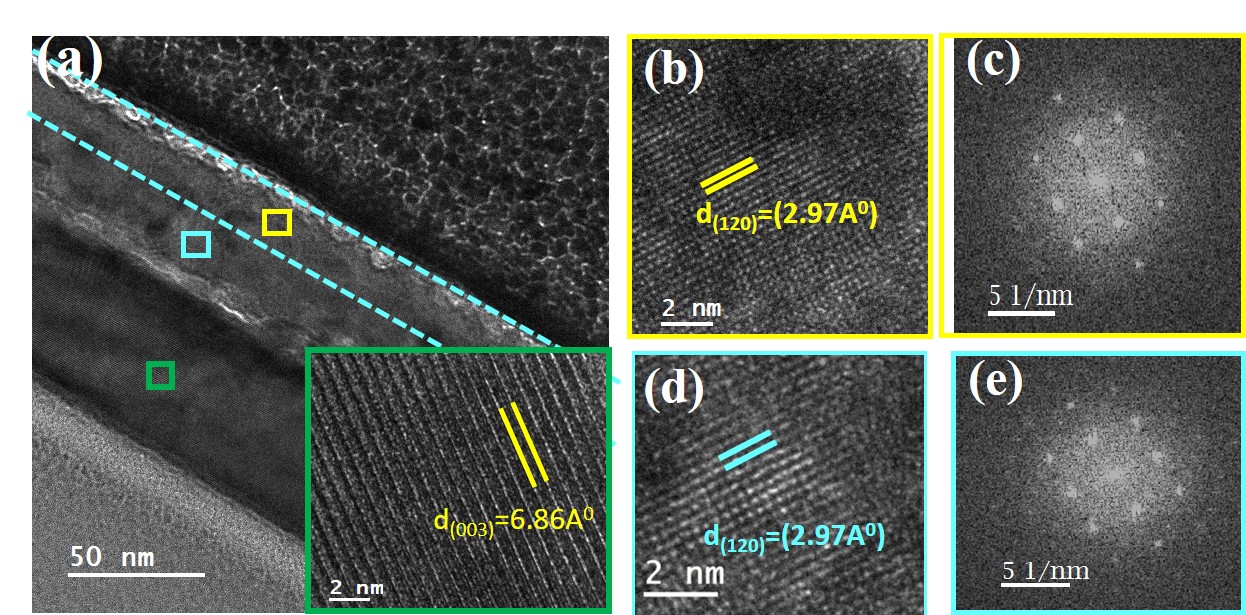}
\caption{The cross-section TEM image of 5$\times10^{13}$ doses of 2D flake: (a) Low-resolution TEM image and inserted image represents the lattice spacing value of 6.86 \AA of the (003) plane without ion implantation near the substrate region; (b) The high-resolution TEM image corresponding to (c) FFT image from the yellow color bounded area in Figure (a) is defined a nearest neighboring lattice plane value 2.97 \AA of the (120) plane. (b) The high-resolution TEM image corresponding (d) FFT image from the clay color bounded area in Figure (e) is defined as the nearest neighboring lattice plane value 2.97 \AA of the (120) plane.}
\label{TEM2}
\end{figure}

The microscopic information about the crystal distortion due to Au ion implantation can be inferred from high-resolution XTEM images. Figure \ref{TEM2} shows the exact thickness and crystal orientation of the low fluence of 5$\times10^{13} ions/cm^2$ implanted 2D CGT flakes. The contrast in the image distinguishes the deposited Pt and the Si/SiO$_2$ substrate. The low magnification image of Figure \ref{TEM2}(a) shows that two staking of CGT flakes are separated from one another with a rough interface. The implanted region is marked with dotted lines. It is expected from the SRIM simulation that the top flake irradiated with Au ions has nearly 20 nm of depth affected by defects. However, low-energy ions are unable to create such promising defects in the system. So, the lower stacking shows a clean fringe pattern throughout the CGT layer, and the inset marked in the green box shows the (003) plane with a lattice spacing of 6.86 \AA $ $. Figure \ref{TEM2}(b) and (d) represent the HRTEM image from the yellow and clay-bounded color area. The lattice spacing value is 2.97 \AA $ $ in the both HRTEM image, corresponding to the (120) lattice plane of CGT. The FFT image of Figure \ref{TEM2}(b) and (d) is defined by the nearest neighbour lattice plane value 2.97 \AA $ $ of the (120) plane. It seems that 5$\times10^{13} ions/cm^2$ does Au ion implantation is unable to significant modification of the crystal structure.

\begin {figure}[!ht]
\centering
\includegraphics[width=3.4in]{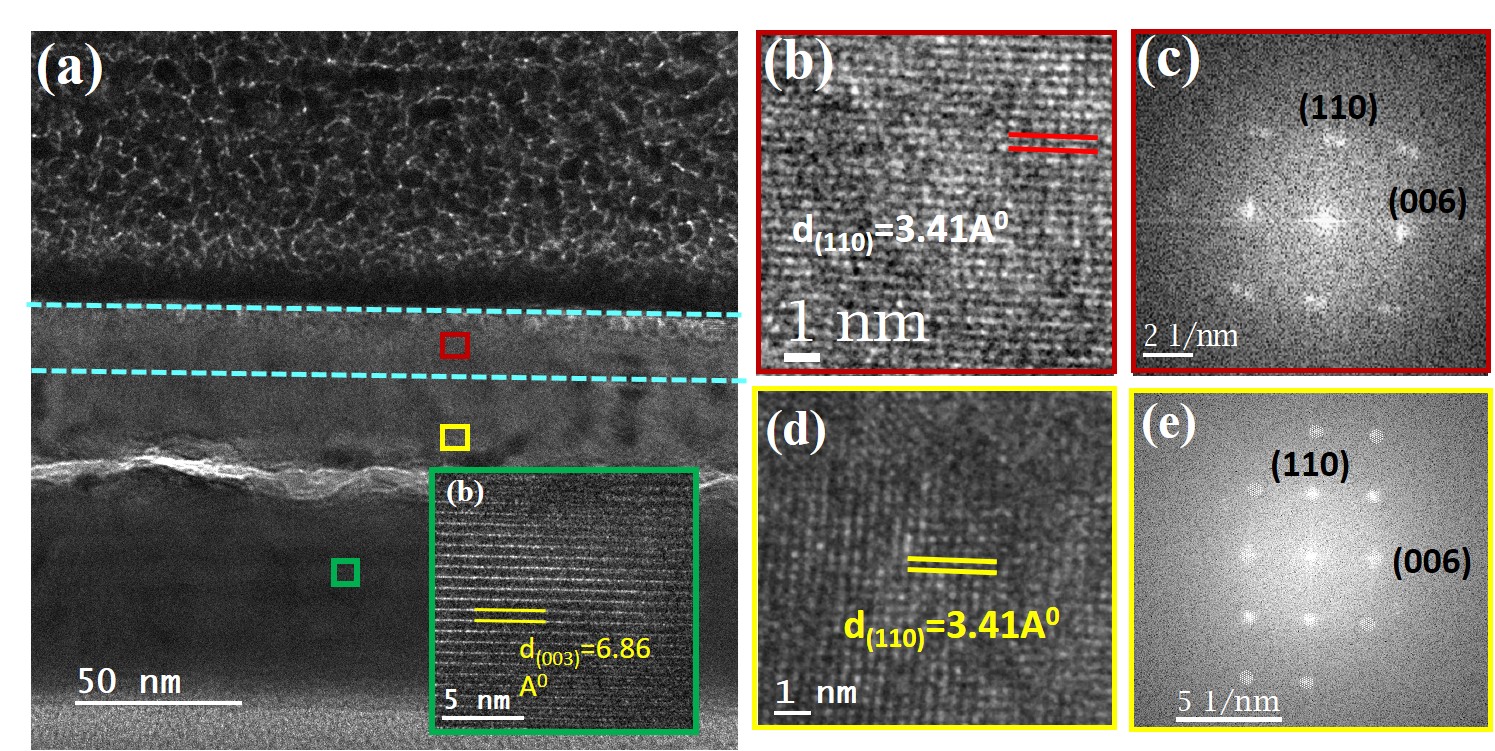}
\caption{The cross-section TEM image of 2.5$\times10^{15} ions/cm^2$ dose 2D flake, (a) Low-resolution TEM image and inserted image represents lattice spacing value of 6.86 \AA $ $ of the (003) plane without ion implantation near the substrate region, (b) The high-resolution TEM image from the red color bounded area in Figure (a). (c) The miler index are define corresponding distorted FFT image. The high-resolution (d) TEM image, (e) the corresponding FFT from the yellow color bounded area in Figure (a) define a nearest neighboring lattice plane value 3.41 \AA $ $ corresponding lattice plane is (110).}
\label{TEM10}
\end{figure}
\begin {figure}[!ht]
\centering
\includegraphics[width=3.4in]{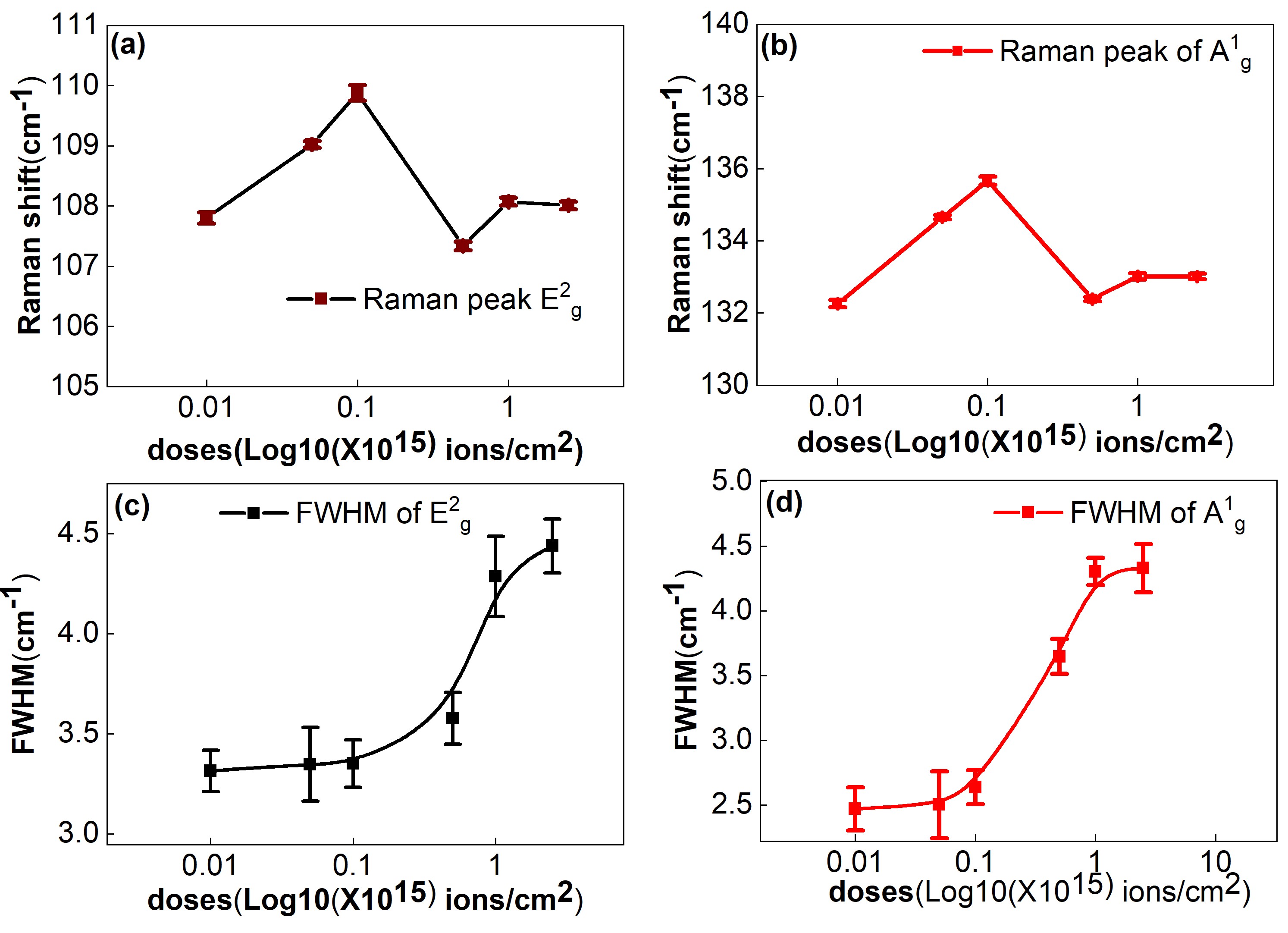}
\caption{ The Raman peaks position of (a) $E^{2}_g$, (b) $A^{1}_g$ modes with the variation of ions fluences.  FWHM of the (c) $E^{2}_g$, (d) $A^{1}_g$.}
\label{Raman1}
\end{figure}
\begin {figure}[!ht]
\centering
\includegraphics[width=3.4in]{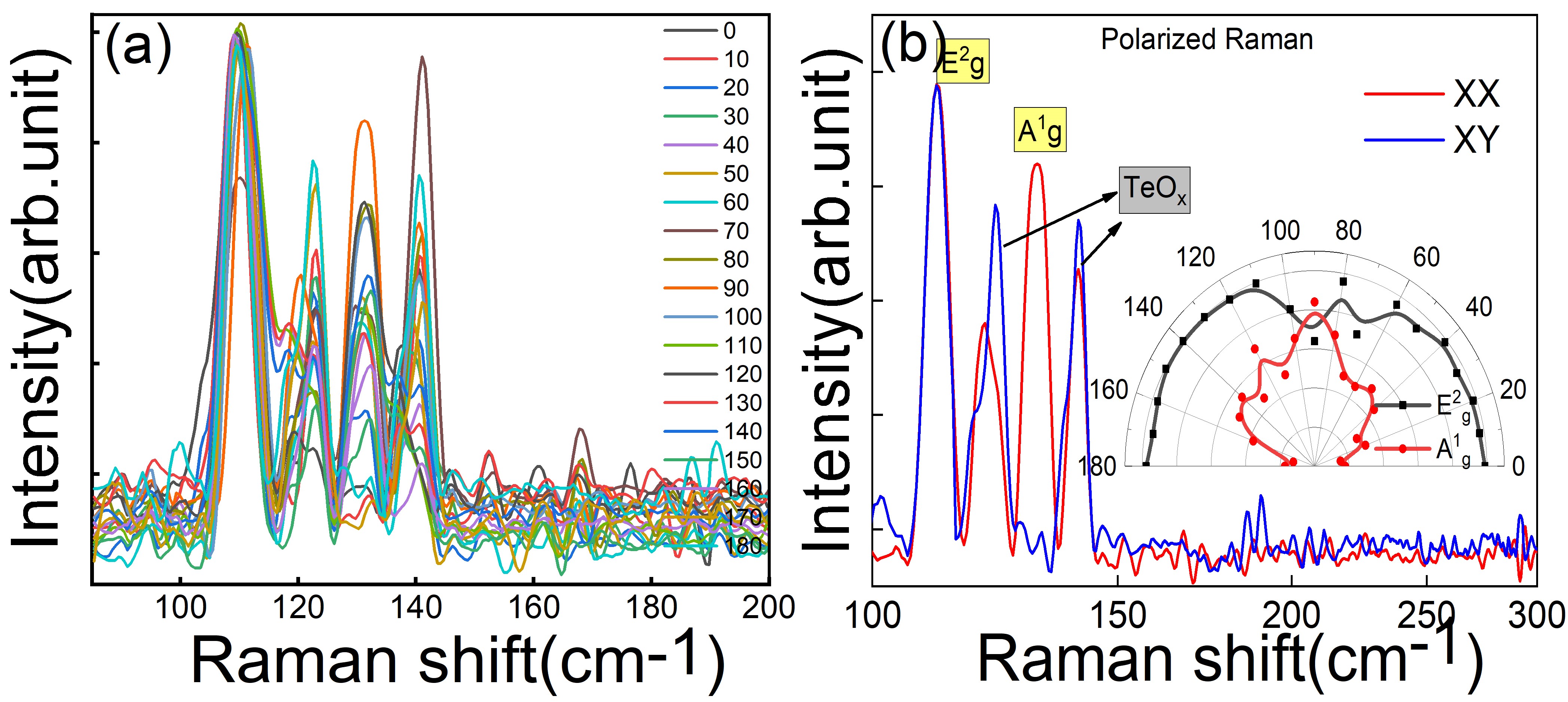}
\caption{(a) The polarisation dependent Raman peak of the CGT flakes in the angle range of 0$^\circ$ to 180$^\circ$. (b) The polarisation dependent Raman peak of the CGT flakes in the angle range of 0$^\circ$ and  180$^\circ$. The inserted figure is the angle-dependent peaks intensity polar plot of the $E^{2}_g$ and $A^{1}_g$.}
\label{POL_Raman}
\end{figure}

Figure \ref{TEM10} exhibits a low-resolution TEM image of a 2D flake implanted with a high dose of 2.5$\times10^{15}ions/cm^2$. The contrast of the image clearly distinguishes the deposited Pt and the Si/SiO$_2$ substrate. The HRTEM and FFT images in Figures \ref{TEM10}(b) and (c) highlight specific regions of interest, i.e. range of implanted with high dose Au ions. In Figure \ref{TEM10}(b), lattice spacing values of 3.41 \AA $ $ are calculated for the (110) lattice planes. But some of cross-sectional part of the c-axis are converted into amorphous patches after ion implantation. Also, the FFT image of Figure \ref{TEM10}(c) shows distorted diffracted spots that signified lattice structural deformation happens for higher doses. The inset of Figure \ref{TEM10}(a) shows the lower stacking of CGT flake with a lattice spacing of 6.86 \AA $ $, representing the (003) plane. The high-resolution TEM and FFT image from the yellow color bounded area of Figure (a), i.e., out-of-range Au ions, are shown in Figure \ref{TEM10}(d) and (e) and the calculated nearest neighboring lattice plane value 3.41 \AA $ $ corresponding lattice plane is (110). Interestingly, the Miller index (006) and (110) of CGT seem to have clean single diffraction spots in Figure \ref{TEM10}(e). This infers that the melting of the surface due to the thermal spike effect reconstructs the surface with different crystal orientations with many point defects at higher doses.  
\begin {figure}[!ht]
\centering
\includegraphics[width=3.4in]{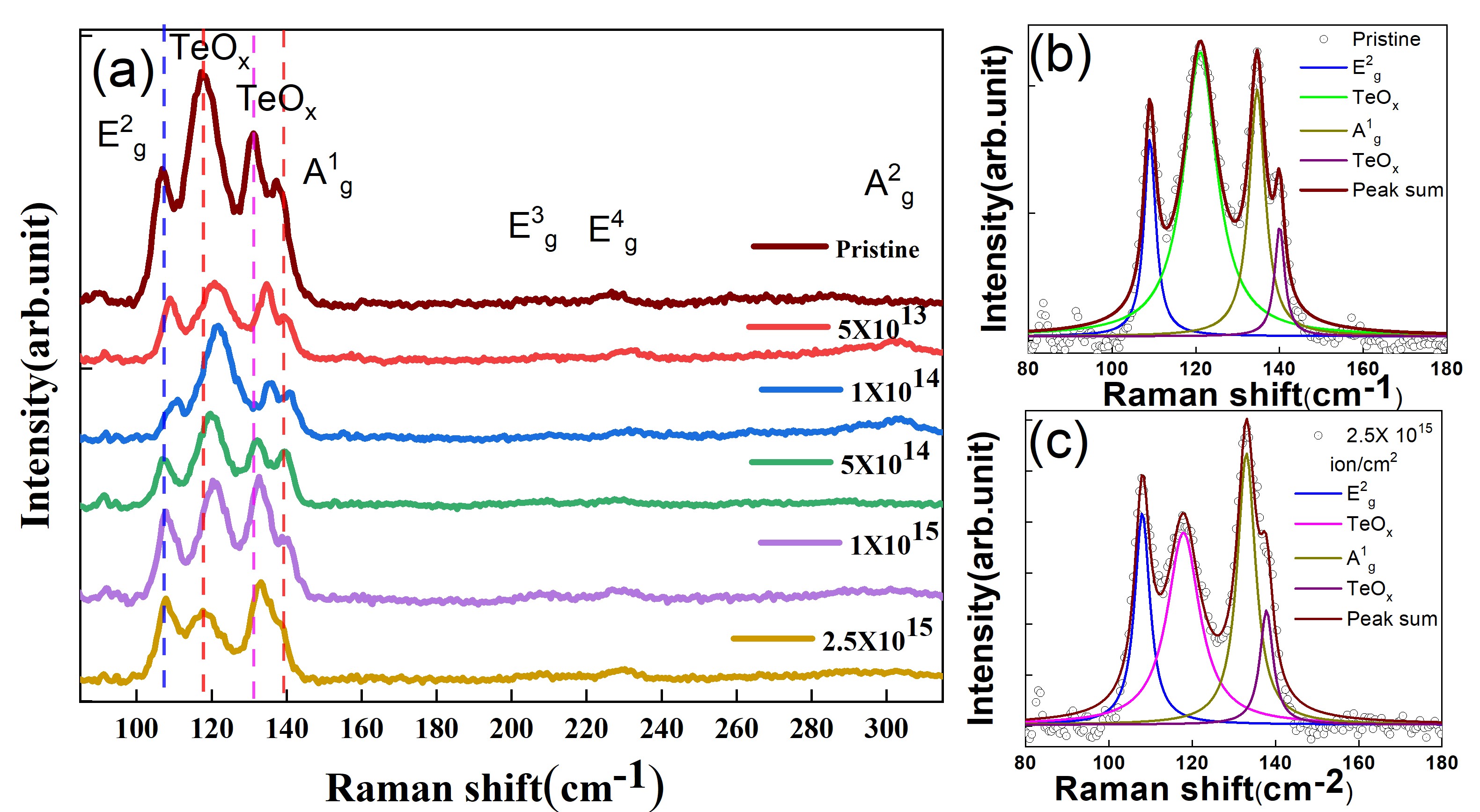}
\caption{ Raman spectra of pristine and implanted samples are obtained, and modes of vibration are identified from the literature survey}
\label{Raman}
\end{figure}
The CGT lattice structure is a 2D van der Waals bonded layer with a central Cr atom surrounded by octahedral Te atoms. According to the theoretical calculation, the expected number of active Raman modes in the hexagonal structure of CGT is ten, but experimentally, only five active modes are observed in the vibrational spectrum. Recently, Lebing Chen et al. reported the long-lived nonlinear optical phonon frequency combs in van der Waals $CrXTe_3$ (X = Ge, Si), revealed by high-resolution Raman spectroscopy, where localized $Ge_2Te_6$ (or $Si_2Te_6$) cluster vibrations generate persistent, coherent, and thermally robust comb structures reproduced by an anharmonic oscillator\cite{Chen2025}. Most of the cases two symmetry Raman active modes (A$_g$, E$^1_{g}$ and E$^2_{g}$) and two time-reversal symmetry protected modes (E$^1_{g}$, E$^2_{g}$) are observed in CGT. However, the surface oxidation suffers to show all Raman modes. The air exposure forms TeO$_x$ due to surface oxidation in CGT, which has a strong Raman peak along with two CGT peaks. Since our samples have several steps after exfoliating to the Si/SiO$_2$ surface and taking to the implantation chamber, etc. So, in our Raman spectra, we have observed TeO$_x$ peaks along with CGT peaks. The Raman spectra of both the pristine and implanted samples are shown in Figure \ref{Raman}(a). The observed peak positions in the Raman spectrum are identified at 110.8 cm$^{-1}$, 136.3 cm$^{-1}$, 212.9 cm$^{-1}$, 233.9 cm$^{-1}$, and 293.8 cm$^{-1}$, corresponding to the $E^{2}_g$, $A^{1}_g$, $E^{3}_g$, $E^{4}_g$, and $A^{2}_g$ modes, respectively, and they match with reported Raman modes\cite{Tian_2016}. Two additional Raman peaks at 120 cm$^{-1}$ and 139 cm$^{-1}$ are observed, corresponding to TeO$_x$ materials, indicating the oxidation of the sample surface\cite{https://doi.org/10.1002/adma.202007792}. In our case, pristine and Au-implanted CGT have two types of experimentally active Raman vibration modes: the $A_g$ mode representing out-of-plane vibrations and the $E_g$ mode representing in-plane vibrations. Among all the Raman modes indicated above, three modes $E^{3}_g$, $E^{4}_g$, and $A^{2}_g$ correspond to Raman frequencies at 212.9 cm$^{-1}$, 233.9 cm$^{-1}$, and 293.8 cm$^{-1}$ are very weak and diminishes with ion fluence. The positions and FWHM of the $E^{2}_g$ and $A^{1}_g$  peaks as a function of ion doses are shown in Figure \ref{Raman1}(a, b, c, d). The frequency shift linearly increases until a fluence of 5$\times10^{14}ions/cm^2$ and then decreases for higher fluence, indicating the softening of the Raman modes is a signature of different thicknesses of each of the CGT materials. Similarly, the FWHM of both peaks initially remains almost unchanged but starts to increase until at higher fluence, indicating the structural deformation after that fluence. This trend is similar to the FWHM of the XRD peaks, confirming that the crystallite size decreases at higher fluences. 
The Raman tensor analysis established that all modes are visible in co-linear (XX) polarized geometry, and $A^{1}_g$ modes vanish in cross-polarization (XY) geometry. We have performed polarization dependence Raman study of CGT to identify the $E^{2}_g$ and $A^{1}_g$ symmetry modes. The details of polarization dependant Raman modes of the CGT flake between polarization angle of 0-180 degree is shown in Figure \ref{POL_Raman}(a), which confirms the presence of in-plane ($E^{2}_g$) peaks and the effective variation of the $A^{1}_g$ mode with angle. Figure \ref{POL_Raman}(b) shows the XX and XY polarized Raman spectra of pristine CGT and found that the 136.3 cm$^{-1}$ corresponds to $A^{1}_g$ symmetry effectively absence in XY polarized mode. The integrated intensity of the polarization-dependent data are plotted in an angular graph between 0-180 degree and shown in the inset of Figure \ref{POL_Raman}(b). The dumbbell shape of angular dependence illustrates the nondegenerate nature of $A^{1}_g$ modes but the intensity of the $E^{2}_g$ are almost unchanged. It also clearly indicates that the intensity of both modes complements each other in the XX and XY polarization geometry. 
From the previous TEM analysis, we have been informed that short-range CGT crystals with amorphous-crystalline patches show within 20 nm thick CGT top layers due to Au ion implantation. We attribute that such distortion crystallinity affects the net magnetization and the direct exchange interaction (DEI) or superexchange interaction (SEI) between (Cr-Te-Cr) coupling. It is well known that charge doping in CGT enhances T$_c$ and affects the FM ordering because of the competition between the AFM direct exchange interaction and the FM superexchange coupling. In order to understand the magnetic phase change properties of Au incorporate CGT, we have performed the magnetization measurement M(T, H) as a function of temperature and applied field for the pristine and implanted samples. 
\begin {figure}[!ht]
\centering
\includegraphics[width=3.4in]{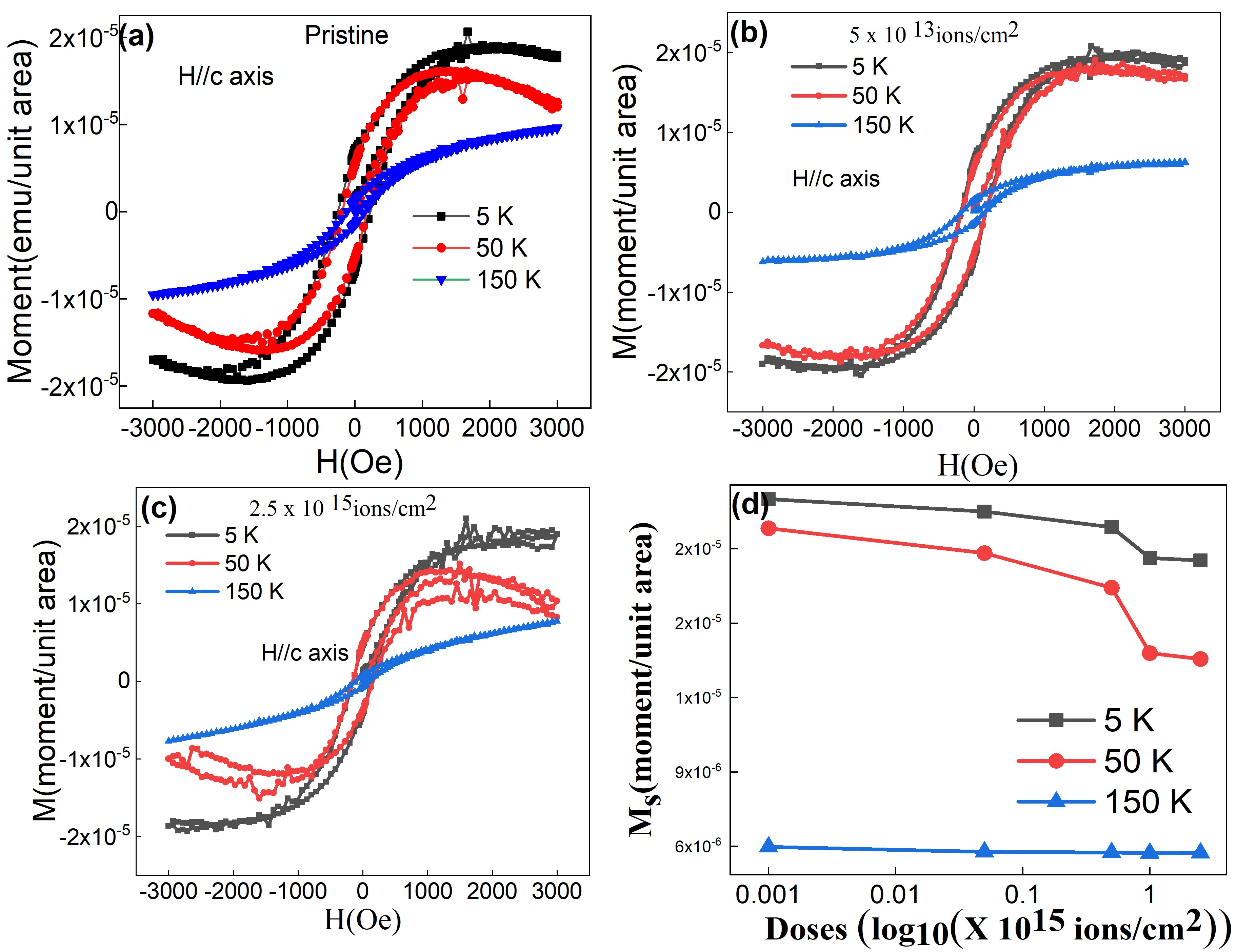}
\caption{MH loop of with the applied field along the c-axis: (a) pristine, (b) 5$\times10^{13} ions/cm^{2}$, (c) 2.5$\times10^{15} ions/cm^{2}$ samples. (d) The saturation magnetisation with the function of ion doses.}
\label{ZFC}
\end{figure}

The M-H curve of the pristine, low-dose (5$\times10^{13} ions/cm^{2}$) and high-dose (2.5$\times10^{15} ions/cm^{2}$) samples, measured with out-of-plane (H$||$ c axis) magnetic field directions, are shown in Figure \ref{ZFC}(a), (b), and (c). For the pristine and low-dose (5$\times10^{15} ions/cm^{2}$) sample, the M-H loop exhibits a prominent hysteresis behavior characteristic of FM materials. The magnetic moment saturates at higher magnetic fields. In the case of the high-dose (2.5$\times10^{15} ions/cm^{2}$) sample, the saturation magnetic moment is reduced compared to the low-dose sample. This reduction can be attributed to the defect-creation sputtering of Te and Ge atoms and a metallic impurity (Au) doping during the ion implantation process. Figure \ref{ZFC}(d) shows the fluence dependant M$_s$ for three different temperatures. From the M-T plot, we know that T$_c$ is around 62 K. So 5K and 50 K show the behavior below T$_c$, and 150 K represents above T$_c$, i.e., in the paramagnetic regions. The value of M$_s$ reduces in the FM state, whereas in the PM regions, it remains unchanged.
\begin {figure}[!ht]
\centering
\includegraphics[width=3.4in]{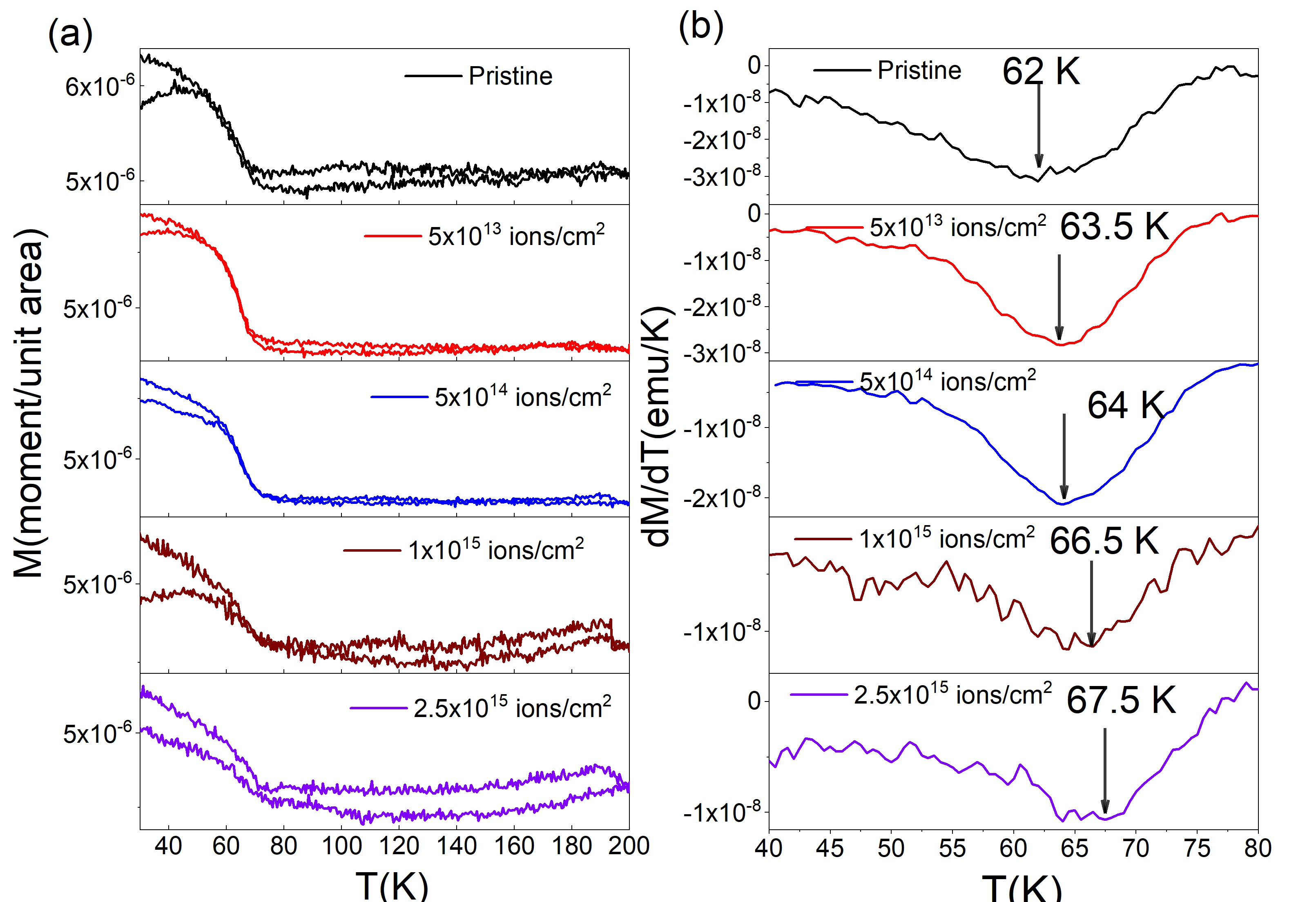}
\caption{(a) ZFC and FC curve with constant applied 300 Oe magnetic field of the pristine, 5$\times10^{13}$, 5$\times10^{14}$, 1$\times10^{15}$ and 2.5$\times10^{15} ions/cm^{2}$ sample. (b) Derivative (dM/dT) of FC curve of the pristine, 5$\times10^{13}$, 5$\times10^{14}$, 1$\times10^{15}$ and 2.5$\times10^{15} ions/cm^{2}$ sample,}
\label{MT}
\end{figure}
Figure \ref{MT}(a) represents the magnetization behavior (M-T) of the pristine and implanted CGT samples. The temperature-dependent ZFC and FC magnetization  (M-T) were performed for out-of-plane (H$||$c axis) along the c-axis in applied fields of 300 Oe from 30 K to 200 K. In the ZFC and FC plots for the pristine and implanted samples (Figure \ref{MT}(a)), the magnetic moment shows an increasing trend below a critical temperature (T$_c$) and continues to increase. This indicates the presence of FM behavior in all implanted samples below T$_c$. It has been observed that the magnetic moment below T$_c$ reduces systematically with ion fluence. This behavior suggests the reduced FM contribution in the sample with ion fluence due to the presence of defects and Au interstitials in the CGT matrix. Figure \ref{MT}(b) shows the derivative of the FC curve for pristine and implanted samples. The transition temperatures are marked for all samples. The magnetic moment as a function of temperature for all samples shows significant changes of the PM to FM transition temperature, and with increasing the ion fluences, the T$_C$ is induced up to 67.5 K for the highest fluence. The M-T data of the out-of-plane applied magnetic field also indicates that the opening of the ZFC and FC curves does not follow the same behavior. In lower doses, the gap between the ZFC and FC curves is larger compared to higher doses, where it is the signature of weak FM signature at higher fluence. The ion implantation process introduces defects and modifies the superexchange interaction, influencing the magnetic properties of the CGT system.
\begin {figure}[!ht]
\centering
\includegraphics[width=3.4in]{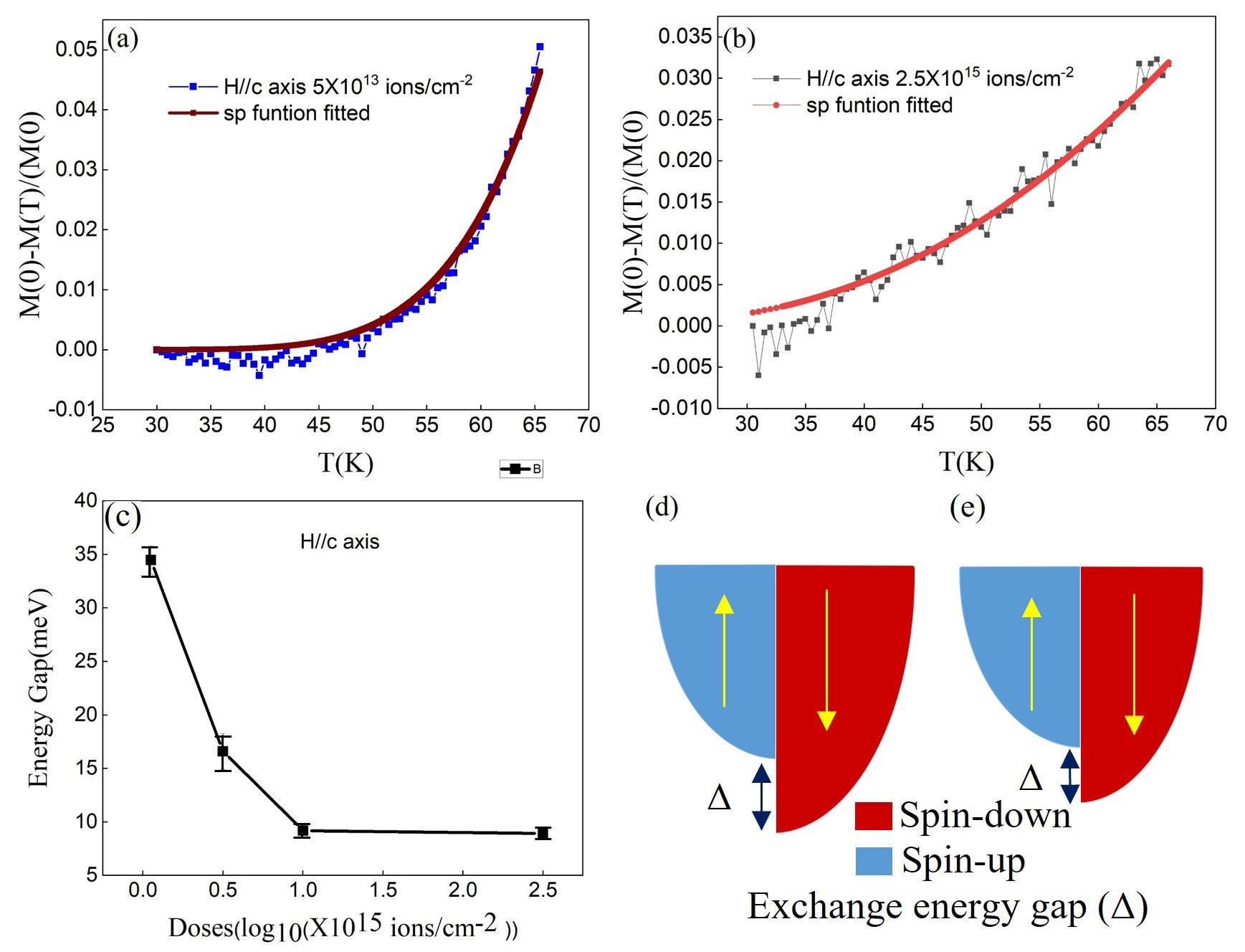}
\caption{Dispersion curve from the FC at H//c axis data of (a) pristine, and (b) 2.5$\times10^{15} ions/cm^{2}$ doses sample and fitted the SP function, (c) The exchange energy gap with the function of ion doses. (d) schematic model of the exchange energy gap for low doses sample. }
\label{band}
\end{figure}

In general, the magnetic moment along the c-axis should saturate below T$_c$, at a relatively low field and is a signature of the alignment of moments in this direction. As shown in Figure \ref{MT}(a), ZFC and FC curves split with the presence of an external magnetic field H = 300 Oe, which originates from the crystalline FM patches, and that varies with ion fluence, i.e., the anomaly below the transition is clearly observed. The decreasing magnetic moments trend may arise due to low-temperature thermal demagnetization. Since our samples have amorphous-crystalline patches, the localized thermal demagnetization can be explained by spin-wave excitations, which follow Bloch's theorem \cite{https://doi.org/10.1016/0304-8853(93)90855-V}. By simplifying the thermal demagnetization due to the excitation of electrons from one subband to the other, the generalized single-particle excitation (SPE) model can be described by the equation\cite{PhysRevB.97.174418}:  

\begin{equation}
    \frac{M(0)-M(T)}{M(0)}=CT^{\frac{3}{2}}exp({- \frac{\Delta}{k_BT}})
\end{equation}
Where M(0) is the magnetic moment at zero temperature, M(T) is the magnetic moment at temperature T, C is a coefficient, $\Delta$ is the energy gap between the top full sub-band and the Fermi level, and $k_B$ is the Boltzmann constant. In the present study, the SPE model was fitted to the field cooling (FC) curves of all the samples, as shown in Figure \ref{band}. By fitting the SPE function to the FC curves in Figure \ref{band}(a), the coefficient values were determined for the pristine sample as follows: C = 0.30$\pm$0.07 and $\Delta$ = 34.47$\pm$1.32.  Figure \ref{band}(b) shows fitting the SP function of the FC curves for the highest doses sample. Figure \ref{band}(c) illustrates the extracted $\Delta$ values obtained from the SPE function fitting using all the FC curves for all doses samples. It is observed that $\Delta$ decreases with increasing ion doses. A larger value of $\Delta$ implies a larger energy gap, indicating the net magnetization reduction at higher doses. A schematic of the reduction of the energy gap between the spin-up and spin-down states via exchange energy is shown in Figure \ref{band}(d) and (e). This effect can be attributed to the involvement of defects and the sputtering of Ge and Te atoms, doping of Au, and surface oxidation, which weaken the strength of the superexchange interaction. The amorphization of CGT induced by low-energy Au$^+$ ion implantation modifies the band structure of the material. Ion implantation causes a decrease in the spin wave (magnon) excitation energy gap, as reflected by the reduction in $\Delta$. Moreover, ion implantation has the ability to modify the anisotropic effects of a single-crystal system. As the ion doses increase more energy is required to orient the magnetic moments of the system in a specific direction. The presence of defects in the lattice caused by ion implantation can lead to a reorientation of some atomic spins within the lattice. This reorientation of spins can contribute to the observed changes in the magnetic behavior of the implanted samples. So, the defect engineering in CGT can tune the magnetic properties that control the energy gap of the system and can be suitable for spintronic applications.

The trigonal crystal structure of CGT is an edge-sharing CrTe$6$ octahedron structure. In this structure, the d orbitals of Cr$^{3+}$ split into two groups: the twofold degenerate E$_g$ and the threefold degenerate t$_{2g}$ orbitals. Three d electrons occupy the t$_{2g}$ orbitals with the same spin orientation, while the E$_g$ orbitals are unoccupied\cite{https://doi.org/10.1002/inf2.12096, coatings12020122}. The bond angle between Cr-3d and Te-3p orbital is approximately 90$^{\circ}$, and the combination of a metal-nonmetal-metal state with a 90$^{\circ}$ bonding network is expected to facilitate superexchange hopping via the nonmetallic atoms, according to the Goodenough-Kanamori rules\cite{Goodenough1955, KANAMORI195987, Anderson1950}. Studies on other Cr-based materials, such as $CrCl_3$, $CrBr_3$, and $CrXTe_3$ (X=Si, Ge, Sb), have demonstrated FM interlayer coupling or weak AFM coupling. Theoretical investigations on $CrSiTe_3$ have indicated that the direct interaction between Cr atoms is short-range, while the superexchange Cr-Te-Cr interaction is a long-range FM interaction dependent on the distance between Cr atoms\cite{PhysRevMaterials.1.014001, Liu2016, Wang2019}. CGT and $Cr_2Si_2Te_6$ have similar crystal structures, the only difference being the substitution of Ge for Si. However, the $T_C$ is doubled in the case of CGT\cite{CHEN201560}. 

\begin {figure}[!ht]
\centering
\includegraphics[width=3in]{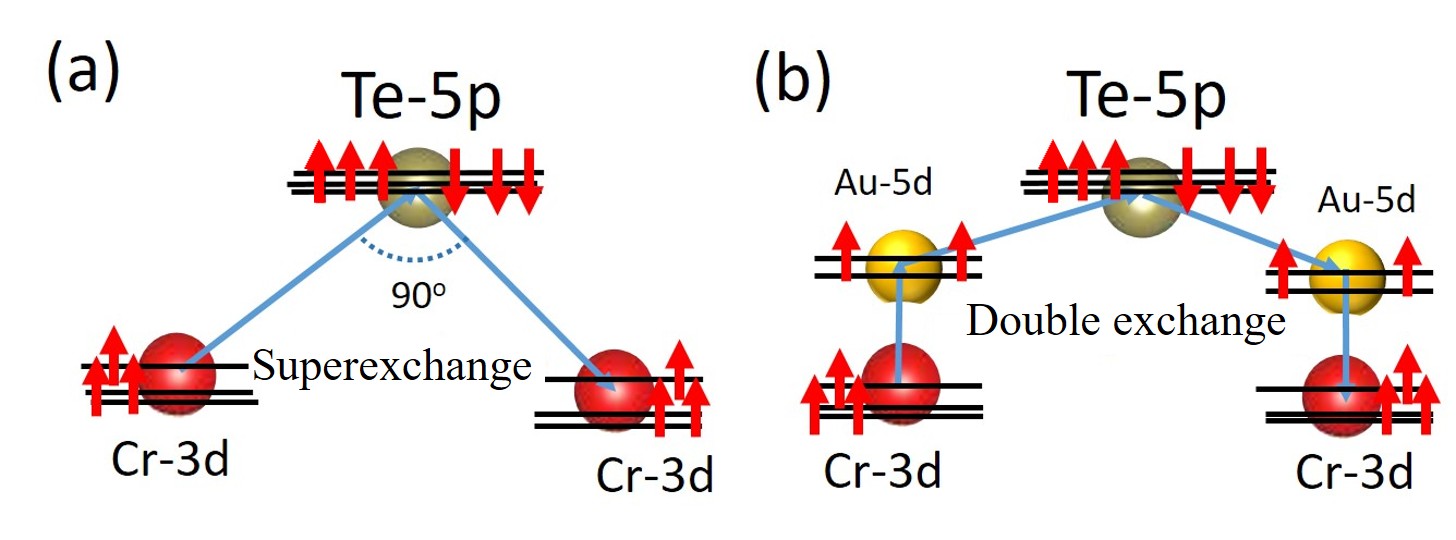}
\caption{(a) Schematic illustration of superexchange coupling between the Cr-Te-Cr 90$^\circ$ bond. (b) The vacancy creation or interstitial Au atoms in between Cr and Te bonds improved the double exchange coupling. }
\label{model1}
\end{figure}
In both cases, Te acts as a ligand facilitating electron hopping. However, in the presence of Ge, a strong covalent bond is formed between Ge dimers and Te 5p orbitals, resulting in occupied and unoccupied bonding and antibonding states. This induces charge transfer from Te 5p to Cr 3d, leading to a decreased effective band gap in the presence of Ge compared to Si atoms\cite{Kim2019}. In the case of ion beam-induced defects, the reduction of out-of plane magnetic moment exhibits a disordered arrangement of magnetic moments within the lattice plane, suggesting the spin-glassy behavior in doped CGT. The defects and vacancies induced due to Au ions effectively reduce the energy gap because of double exchange between Cr and Te via Au - 3d states\cite{10.1021/jacs.9b06929}. However, in our case, the $T_C$ induced after Au doping. The origin of the $T_C$ increasing with the fluences is expected to be the double exchange interaction induced via Au interstitial. Here, Figure \ref{model1}(a) is the model of the superexchange in Cr(3d)-Cr(3d) via ligand Te(5p). The Figure \ref{model1}(b) represents the double exchange interaction between Cr(3d)-Cr(3d) via Au(5d) and ligand Te(5p). Here, Au doping reacts as a donor spin during double exchange interaction, and this signature indicates the origin of induced the $T_C$ for high doses. The defects will be more for higher doses of ion implantation which is investigated using XRD, Raman and TEM characterisation.  The induced defect in CGT materials has reduced the overall magnetic moment presence in the system. But the Au ions interstitial position will originate another fundamental double exchange interaction.  This ion implantation study in CGT has a signature for inducing the Curie temperature ($T_C$) of 2D magnetic materials.

\section{Conclusions}
 The 2D FM CGT crystal was synthesized using multistep annealing for a few days. We successfully exfoliated 2D single crystal CGT onto the Si/SiO$_2$ using Scotch tape methods. Low energy (30keV) Au ions were implanted at different fluences on the 2D flakes of CGT to create defects in the lattice. The FESEM and EDX confirm the effect of the thermal spike on the surface of CGT at higher fluences and reduction of Te, Ge, and Cr atomic percentage and the increment of Au atoms due to sputtering. XTEM images concluded that the ions damaged the lattice surface and van der Waals structure. However, the FWHM of the XRD illustrates the role of lattice defects up to a certain fluence and decreases the crystallinity due to ion beam-induced heating. The mode softening at the same ion fluence matches well with the XRD data. The low doses have not created more defects, so the FM properties of the 2D CGT are preserved. As per increasing the higher amounts of ions, the FM properties disintegrate and induce an intermediate path for exchange interactions that are reflected from the out-of-plane applied field (300 Oe) magnetization measurement. The exchange energy gap calculated from the FC curve decreases with ions fluence, indicating the energy gap between spin-up and spin-down states reduces. The defect engineering by ion implantation in CGT can tune the magnetic properties and control the exchange interaction energy gap of the system. Such phenomena can be useful in ferromagnetic spintronic device applications.
\section{acknowledgments}
The authors would like to thank the National Institute of
Science Education and Research (NISER), Department of Atomic Energy (DAE), Government of India for funding the research work through the project number RIN-4001. The authors acknowledge the staff of low energy ion beam facility of the Institute of Physics (IOP), Bhubaneswar for providing stable beams during ion implantation. PKS and GG thank Dr. Kartik Senapati for fruitful scientific discussion.

\nocite{*}
\bibliography{aipsamp}% Produces the bibliography via BibTeX.

\end{document}